# Global gyrokinetic simulations of the impact of magnetic island on ion temperature gradient driven turbulence

J. C. Li[1,6*], J. Q. Xu[2,6], Y. R. Qu[3], Z. Lin[4*], J. Q. Dong[2,5], X. D. Peng[2], J. Q. Li[2]

[1] Department of Earth and Space Sciences, Southern University of Science and Technology, Shenzhen, Guangdong 518055, China
[2] Southwestern Institute of Physics, PO Box 432, Chengdu 610041, China
[3] School of Physics, Nankai University, Tianjin 300071, China
[4] University of California, Irvine, CA 92697, United States of America
[5] ENN Science and Technology Development Co., Ltd., Langfang 065001, China
[6] Joint first authors
Email: jingchunli@pku.edu.cn
       zhihongl@uci.edu

**Abstract:** The effect of island width on the multi-scale interactions between magnetic island (MI) and ion temperature gradient (ITG) turbulence has been investigated based on the global gyrokinetic approach. It is found that the coupling between the island and turbulence is enhanced when the MI width ($w$) becomes larger. A vortex flow that is highly sensitive to the width of the magnetic island can be triggered, ultimately resulting in a potent $E \times B$ shear flow and a consequent reduction in turbulent transport. The shearing rate induced by the vortex flow is minimum at the O-point while it is maximum at both of the two reconnection points of the island, i.e., the X-points, regardless of the island width. There exists a nonmonotonic relationship between zonal flow (ZF) amplitude and island width, showing that the ZF is partially suppressed by medium-sized MIs whereas enhanced in the case of large island. A larger MI can tremendously damage the ITG mode structure, resulting in higher turbulent transport at the X-point whereas a lower one at the O-point, respectively. Such phenomenon will be less distinct at very small island widths below $w/a \sim 8\% \approx 12\,\rho_i$ ($a$ is the minor radius and $\rho_i$ the ion gyroradius), where it shows that turbulence near the X-point is hardly affected although it is still suppressed inside the island. Furthermore, the influence of different island sizes on turbulence transport level is also discussed.

# 1. Introduction

Understanding the physics of interactions between magnetic island (MI) and turbulence is a crucial topic in magnetically confined plasma research which has attracted much attention recently [1-3]. On one hand, magnetic island [4] is a ubiquitous magnetohydrodynamic (MHD) structure induced by magnetic reconnection [5] in both magnetized and space plasmas and associated physics has been extensively studied, particularly in the field of controlling techniques of tearing mode (TM) [6] or neoclassical tearing mode (NTM) [7] which would restrict the high performance discharges of tokamaks. On the other hand, turbulence regulates the energy flux of plasma from small-scale instability to large-scale flow, which in turn affects global confinement performance [8]. Besides, it is well known that drift wave instabilities driven by steep density and temperature gradients such as ion temperature gradient mode (ITG) [9-11] and trapped electron mode (TEM) [12] can cause anomalous transport thus controlling the turbulence is a major task in the present-day fusion research. The drift-wave turbulence fluctuations can self-organize into a linearly stable zonal flow (ZF) in the nonlinearly saturated stage, which acts as an energy sink for turbulence [13-15]. Physically, the macro-scale MHD activities are closely related to zonal flows and micro-turbulence. Any coupling to magnetic island opens an additional channel for turbulent energy dissipation and thus could provide an important mechanism for stabilization of turbulence. Hence, investigating the interaction between the MI and plasma turbulence could not only promote the understanding of the macro-instability but also provide potential controlling mechanisms of plasma performance in future fusion devices such as ITER.

Over the last decade, much effort has been made in identifying the background microturbulence in the presence of magnetic islands. Recent studies have primarily centered around the impact of magnetic islands (MIs) on ion-scale microturbulence in tokamak plasmas. These investigations have revealed that the interplay between multi-scale phenomena occurs through mechanisms such as equilibrium modulation or the presence of meso-scale structures like zonal flows (ZFs)[16]. For example, in the confinement devices such as the J-TEXT tokamak and the LHD stellarator, it is observed that when the magnetic island exists, there is will be a strong vortex $E \times B$ shear flow, which triggers the suppression of turbulence by magnetic islands [17, 18]. At the magnetic island separatrix, it is generally believed that shear flow is enhanced. The first measurements of localized reduction of turbulent density fluctuations as well as the reduction of cross-field transport at the O-point of $m/n$=2/1 NTMs is reported for DIII-D experiments ($m$ and $n$ are poloidal and toroidal mode numbers, respectively) [19]. Turbulence and its modulation by the flow shear development around the MI in KSTAR is clarified [20] and a quadruple vortex flow was demonstrated in the J-TEXT experiment when the island width is small [21]. In addition, experiments on the HL-2A tokamak have shown that the modulation effect of the magnetic island on turbulence is limited to the inner region of the magnetic island and there is a magnetic island threshold for modulating the turbulence around $w$~4 cm [22], which is consistent with Fitzpatrick's theoretical predictions [23]. Most recently, the multi-scale interactions among macro-scale tearing mode, meso-scale geodesic acoustic mode (GAM) and small-scale turbulence are carried out in the edge plasma of the HL-2A tokamak using the causality analysis method and the effect of the magnetic island width on the nonlinear coupling among turbulence, GAM and tearing modes is also reported [24, 25].

On the theoretical side, H. R. Wilson *et al* established a set of ITG eigen equations describing the existence of magnetic islands based on the shear slab geometry, thus explaining the mechanism of the linear ITG mode being modulated in the island [26]. P. Jiang *et al* also verified the modulation effect of magnetic islands on ITG in the linear phase [27] employing the global gyrokinetic code (GTC) [28]. In addition to the modulation effect on the ITG mode in the linear phase, sufficiently large magnetic islands also suppress the turbulence level in the nonlinear phase. The influence of rotating islands on ITG turbulence has been examined by Zarzoso *et al* using gyrokinetic code GKW, which is suggested that in the nonlinear phase the MIs can suppress the turbulence level inside the islands [29]. The turbulence level inside the island can be reduced up to 50% in comparison with the case of in the absence of MIs. Similar results has also been confirmed by local GENE simulations which shows that the helical flow with different shear intensities are increased on both sides of the magnetic island [30]. In addition to the suppression of drift wave instability and turbulence, several mode coupling effects caused by magnetic islands have also been found in numerical simulations, for instance, fluid simulation by Wang *et al* found that when the MI width exceeds a certain level, a new radial non-localized ITG mode will be excited inside its region, which is called the magnetic island induced ITG mode [31]. W. A. Hornsby *et al* also found that the MI can resonantly couple to the low-*n* mode generated in the nonlinear stage of ITG turbulence and excite a vortex-like large scale potential structure, i.e. vortex flow [32]. Besides, a portion of the influence of turbulence on magnetic islands can be attributed to the polarization current effects, which arises from the acceleration of flow around the separatrix of the magnetic island. The polarization current effects have been discussed in previous literature. For instance, J. W. Connor et al. conducted a two-fluid, cold ion, collisional analysis to investigate the role of the polarization current in the evolution of magnetic islands in slab geometry [39]. They specifically examined the combined effect of parallel electron dynamics and perpendicular transport (particle diffusion and viscosity) on the distribution of the polarization current within the magnetic islands. By employing a method that involves merging small-scale islands to eliminate the nonlinear excitation of coherent magnetic islands by turbulence, researchers were able to specifically isolate and study the effects of the polarization current. The findings revealed that the polarization current induced by turbulence can indeed have a destabilizing effect on magnetic islands, thereby impacting their overall stability [40, 41]. Furthermore, the influence of the polarization current on the propagation of magnetic islands has also been studied. It is observed that for larger islands with widths significantly exceeding the ion gyroradius ($w \gg \rho_i$), the free propagation velocity of the islands matches the velocity of the E × B flow [42, 43]. On the contrary, for smaller islands with widths comparable to the ion gyroradius ($w \approx \rho_i$), the propagation is significantly influenced by the ion diamagnetic flow. Although numerous experimental observations and simulations have suggested the influence of MI on turbulence and transport, the understanding of the interaction between MIs and turbulence is not clear enough, mainly due to its multi-scale, kinetic characteristic and highly nonlinear physical nature, which makes it difficult to describe by simple theoretical models. The in-depth explanation of the nonlinear interaction between MIs, self-generated vortex flows and turbulence are of great importance in promoting the understanding of the physics of multi-scale interactions, which is the primary goal of the present paper.

In this work, we present the global gyrokinetic analysis of the impact of magnetic island size on turbulent transport as well as the nonlinear multi-scale interactions. The rest of the paper is organized as follows. The physical model is presented in section 2. The simulation results of the impact of island size on vortex flow, local turbulence and transport are demonstrated in section 3. Finally, the conclusions are drawn in section 4.

## 2. Physical model

The simulation model used in the present study is based on the following gyrokinetic equation [38, 39]

$$\frac{d}{dt}f_s(\boldsymbol{R}, v_{||}, \mu, t) = \left(\frac{\partial}{\partial t} + \dot{\boldsymbol{R}} \cdot \nabla + \dot{v}_{||}\frac{\partial}{\partial v_{||}} - C_{coll}\right)f_s = 0. \quad (1)$$

where $f_s(\boldsymbol{R}, v_{||}, \mu, t)$ is the distribution function for each particle species $s$ described in five dimensional spaces: the gyrocenter position $\boldsymbol{R}$, the parallel velocity $v_{||}$, and the magnetic momentum $\mu$. $\dot{\boldsymbol{R}}$ is the gyrocenter velocity, $\dot{v}_{||}$ is the parallel acceleration and $C_s$ is the collision operator, respectively. Collisions are neglected in the simulations, namely $C_{coll} = 0$. In the framework of gyrokinetic theory, $\dot{\boldsymbol{R}}$ and $\dot{v}_{||}$ can be expressed as:

$$\dot{R} = v_\| \frac{B}{B_\|^*} + v_E + v_c + v_g, \quad (2)$$

$$v_\|^* = -\frac{1}{m_s}\frac{B^*}{B_\|^*} \cdot (\mu \nabla B + Z_s \nabla \delta\phi), \quad (3)$$

where $Z_s, m_s, c, \boldsymbol{B}, \delta\phi$ are the electric charge, the particle mass, the light speed, the background magnetic field and the perturbed electrostatic potential, respectively. The expression $\boldsymbol{b} \equiv \frac{\boldsymbol{B}}{B}$ represents the unit magnetic vector. $B^* = B + \frac{m_s c}{Z_s} v_\| \nabla \times b$. $B_\|^* = B^* \cdot b$ corresponds to the Jacobian determinant in phase space. The gyrocenter velocity is consist of the parallel velocity $v_\| \frac{B}{B_\|^*}$, the $E \times B$ drift velocity $v_E = \frac{cb \times \nabla \delta\phi}{B_\|^*}$, the magnetic curvature drift velocity $v_c = \frac{m_s c}{Z_s B_\|^*} v_\|^2 \nabla \times b$ and the magnetic gradient drift velocity $v_g = \frac{c\mu}{Z_s B_\|^*} b \times \nabla B$.

The implementation of the magnetic islands in GTC code [28] is achieved by adding a extra perturbation on the background magnetic field, that is, $\boldsymbol{B} = \boldsymbol{B_0} + \delta\boldsymbol{B_I}$. References [27, 40] have extensively discussed the implementation of this process in the GTC code. For the sake of completeness, we also provide here the procedure for introducing static magnetic islands within the gyrokinetic framework. In general, the magnetic field caused by the MIs is much weaker compared to the background magnetic field with a typical order about $\frac{|\delta B_I|}{|B_0|} \sim 10^{-4}$, thus the following approximations are utilized [40]:

$$\begin{cases} \boldsymbol{v_E} = \dfrac{c(\boldsymbol{b_0} + \delta\boldsymbol{B_I}/B_0) \times \nabla\delta\phi}{B_\|^*} \\ \boldsymbol{v_c} = \dfrac{m_s c}{Z_s B_\|^*} v_\|^2 \nabla \times \boldsymbol{b_0} \\ \boldsymbol{v_g} = \dfrac{c\mu}{Z_s B_\|^*} \boldsymbol{b_0} \times \nabla B_0 \\ \nabla \times \boldsymbol{b} \cdot \nabla B = \nabla \times \boldsymbol{b_0} \cdot \nabla B_0 \\ \nabla \times \boldsymbol{b} \cdot \nabla \delta\phi = \nabla \times (\boldsymbol{b_0} + \delta\boldsymbol{B_I}/B_0) \cdot \nabla\delta\phi. \end{cases} \quad (4)$$

The above expressions imply that only the effect of the MIs on the $E \times B$ drift velocity is retained since it plays a more important role in the ITG instability and the generation of the helical shear flow than the other two terms. We also keep $B_\|^*$ to its first order as can be written as follows:
$$\begin{cases} B_\|^* = B_{0\|}^* + \delta B_\|^* \\ B_{0\|}^* = \left(\boldsymbol{B_0} + \frac{m_s c}{Z_s} v_\| \nabla \times \boldsymbol{b_0}\right) \cdot \boldsymbol{b_0} \\ \delta B_\|^* = 2\delta \boldsymbol{B_I} \cdot \boldsymbol{b_0} + \frac{m_s c}{Z_s} v_\| \left[\nabla \times \left(\frac{\delta \boldsymbol{B_I}}{B_0}\right) \cdot \boldsymbol{b_0} + \nabla \times \boldsymbol{b_0} \cdot \frac{\delta \boldsymbol{B_I}}{B_0}\right]. \end{cases} \quad (5)$$

The expressions of $\dot{\boldsymbol{R}}$ and $\dot{v}_\|$ can be approximated as:

$$\dot{\boldsymbol{R}} = v_\| \frac{\boldsymbol{B_0} + \delta \boldsymbol{B_I}}{B_\|^*} + \frac{c\left(\boldsymbol{b_0} + \frac{\delta \boldsymbol{B_I}}{B_0}\right) \times \nabla \delta \phi}{B_\|^*}$$
$$+ \frac{m_s c}{Z_s B_\|^*} v_\|^2 \nabla \times \boldsymbol{b_0} + \frac{c\mu}{Z_s B_\|^*} \boldsymbol{b_0} \times \nabla B_0, \quad (6)$$

$$\dot{v}_\| = -\frac{1}{m_s} \frac{\boldsymbol{B_0^*} + \delta \boldsymbol{B_I}}{B_\|^*} \cdot (\mu \nabla B_0 + Z_s \nabla \delta \phi) - cv_\| \frac{\nabla \times \left(\frac{\delta \boldsymbol{B_I}}{B_0}\right)}{B_\|^*} \cdot \nabla \delta \phi. \quad (7)$$

According to the principle of the $\delta f$ method, the propagator $d/dt \equiv L$ and the distribution function $f_s$ need to be split into an equilibrium part without magnetic islands and a perturbed part. First let the propagator $L = L_0 + \delta L$, we have:

$$L_0 = \frac{\partial}{\partial t} - \frac{\mu}{m_s} \frac{\boldsymbol{B_0^*}}{B_{0\|}^*} \cdot \mu \nabla B_0 \frac{\partial}{\partial v_\|} + v_\| \frac{\boldsymbol{B_0}}{B_{0\|}^*} \cdot \nabla$$
$$+ \left(\frac{m_s c}{Z_s B_\|^*} v_{0\|}^2 \nabla \times \boldsymbol{b_0} + \frac{c\mu}{Z_s B_{0\|}^*} \boldsymbol{b_0} \times \nabla B_0\right) \cdot \nabla, \quad (8)$$

$$\delta L = \left(\frac{B_{0\|}^*}{B_\|^*} - 1\right) L_0 + \left(v_\| \frac{\delta \boldsymbol{B_I}}{B_\|^*} + \frac{c\left(\boldsymbol{b_0} + \frac{\delta \boldsymbol{B_I}}{B_0}\right) \times \nabla \delta \phi}{B_\|^*}\right) \cdot \nabla$$
$$- \frac{1}{m_s} \frac{\boldsymbol{B_0} + \delta \boldsymbol{B_I}}{B_\|^*} \cdot Z_s \nabla \delta \phi \frac{\partial}{\partial v_\|} - \frac{1}{m_s} \frac{\delta \boldsymbol{B_I}}{B_\|^*} \cdot \mu \nabla B_0 \frac{\partial}{\partial v_\|}$$
$$- cv_\| \frac{\nabla \times \left(\boldsymbol{b_0} + \frac{\delta \boldsymbol{B_I}}{B_0}\right)}{B_\|^*} \cdot \nabla \delta \phi \frac{\partial}{\partial v_\|}. \quad (9)$$

Similarly, we split the distribution function $f_s = f_{0s} + \delta f_s$, where $f_{0s}$ is the equilibrium part which is time-independent satisfying $L_0 f_{0s} = 0$. The solution of this equation is

$$f_{0s} = \frac{n_{0s}}{(2\pi v_{th,s}^2)^{1.5}} \exp\left(-\frac{E_s}{T_{0s}}\right), \quad (10)$$

which is a local Maxwellian. $n_{0s}, T_{0s}$ are the initial equilibrium density and temperature of the system, $v_{th,s} = \sqrt{T_{0s}/m_s}$ is the thermal speed, and $E_s = \frac{1}{2}m_s v_\parallel^2 + \mu B_0$ is the kinetic energy of the particle. Defining the particle weight as $w_s = \delta f_s/f_s$ and considering that $L_0 f_{0s} = 0$ and $(L_0 + \delta L)(f_{0s} + \delta f_s) = 0$, we can obtain the evolution equation for $w_s$:

$$\frac{d}{dt} w_s = -\frac{1}{f_s}\delta L f_{0s} =$$
$$-(1 - w_s) \times \begin{bmatrix} \underbrace{\frac{c(\boldsymbol{B_0} + \delta \boldsymbol{B_{IS}}) \times \nabla \delta \phi}{B_0^2} \cdot \frac{1}{f_{0s}} \nabla \Big|_{v_\perp} f_{0s}}_{E \times B} \\ + \underbrace{\frac{v_\parallel Z_s}{T_{0s}}\left(\frac{\boldsymbol{B_0} + \delta \boldsymbol{B_{IS}}}{B_0} \cdot \nabla \delta \phi\right)}_{parallel} \\ + \underbrace{\frac{Z_s}{T_{0s}}\left(\frac{v_\parallel^2}{\Omega_s}\nabla \times \boldsymbol{b_0} + \frac{\mu}{m_s \Omega_s}\boldsymbol{b_0} \times \nabla B_0\right) \cdot \nabla \delta \phi}_{magnetic\ drift} \\ + \underbrace{v_\parallel \frac{\delta \boldsymbol{B_{IS}}}{B_0} \cdot \frac{1}{f_{0s}} \nabla \Big|_{v_\perp} f_{0s}}_{flatten\ effect} \end{bmatrix}. \quad (11)$$

Here $\Omega_s = Z_s B_0/m_s c$ is the particle cyclotron frequency with the gradient operator $\nabla \Big|_{v_\perp} f_{0s} = \left(\nabla + \frac{\mu \nabla B_0}{T_{0s}}\right) f_{0s}$. As the effects of magnetic islands on drift terms are ignored hence we have let $B_{0\parallel}^* = B_\parallel^* = B_0$.

Once $\delta f_s$ is obtained, the perturbation density, current, and pressure can be obtained from the integration of the function in velocity space:

$$\begin{cases} \delta n_s(\boldsymbol{x}, t) = \int_{R \to x} d\boldsymbol{v}\, \delta f_s(\boldsymbol{R}, \mu, v_\parallel, t), \\ \delta J_{\parallel s}(\boldsymbol{x}, t) = \int_{R \to x} d\boldsymbol{v}\, v_\parallel Z_s \delta f_s(\boldsymbol{R}, \mu, v_\parallel, t), \\ \delta P_{\parallel s}(\boldsymbol{x}, t) = \int_{R \to x} d\boldsymbol{v}\, m_s v_\parallel^2 \delta f_s(\boldsymbol{R}, \mu, v_\parallel, t), \\ \delta P_{\perp s}(\boldsymbol{x}, t) = \int_{R \to x} d\boldsymbol{v}\, \mu B_0 \delta f_s(\boldsymbol{R}, \mu, v_\parallel, t). \end{cases} \quad (12)$$

Here the integral operator $\int_{R \to x} d\boldsymbol{v} \equiv \int \frac{2\pi B_0}{m_s} d\mu dv_\parallel \int \frac{d\alpha}{2\pi} d\boldsymbol{R}\, \delta(\boldsymbol{R} + \boldsymbol{\rho_s} - \boldsymbol{x})$ is used to convert values from the gyrocenter coordinate back to the real coordinate.

Besides, the gyrokinetic Poisson equation is

$$\frac{Z_i^2 n_{0i}}{T_{0i}}(\delta\phi - \delta\widetilde{\phi}) = Z_i \delta n_i + Z_e \delta n_e. \quad (13)$$

Here $\delta\widetilde{\phi}(\mathbf{x},t) = \frac{1}{n_i}\int_{\mathbf{R}\to\mathbf{x}} d\mathbf{v}\, f_i(\mathbf{R},\mu,v_{||},t)\delta\overline{\phi}(\mathbf{R},t)$ is the result of transforming the cyclotron-averaged potential $\delta\overline{\phi}(\mathbf{R},t)$ back to the real coordinates.

Equations (1-13) form a closed system and can be utilized to simulate the ITG instability in the plasma with magnetic islands. For the electron, the gyro-radius $\rho_e$ is small enough whose Larmor radius effect can be neglected, and under such assumptions we obtain the drift kinetic electron model (DKE). While the hybrid model addresses the zonal and nonzonal components separately, the DKE model tackles both components simultaneously, which is necessary when magnetic islands are present. This approach is essential to accurately capture the complex dynamics of plasma turbulence in tokamaks, where the presence of magnetic islands can significantly affect the behavior of both zonal and nonzonal parts. This is in contrast to the hybrid model, which does not fully account for the impact of magnetic islands on the plasma behavior. The DKE model therefore represents an important advance in our understanding of plasma turbulence and its role in tokamak performance.

## 3. Simulation results

### 3.1. Simulation setup

The Cyclone base case-like equilibrium [41] is chosen in our simulations, the equilibrium safety factor ($q$) profile and the temperature and density profile profiles utilized in the simulations is shown in Fig. 1. Here we focus on the effect of $m/n=2/1$ magnetic island on the nonlinear evolution of bulk plasma turbulence. At the plasma axis, the ion and electron temperature are $T_e=T_i=2.22$ keV, the densities are $n_e = n_i = 1.13 \times 10^{13} cm^{-3}$, the magnetic field strength is $B_0=2.01254\times 10^4$ G, the dynamical plasma beta $\beta_e = 8\pi n_e T_e/B_0^2 = 0.25\%$, and the ion gyroradius $\rho_i/R_0 = m_i c\sqrt{kT_i/m_i}/(BeR_0) = 2.86 \times 10^{-3}$, where $m_i$ is the ions' mass, $k$ is Boltzmann's constant. The $q=2$ rational surface (RS) is located at the center of the plasma $r=0.5a$, where the magnetic shear $s=(r/q)(dq/dr)=0.54$. Furthermore, the characteristic scale lengths of particle density and temperature are defined as $L_n=-(d\ln n/dr)^{-1}$, and $L_T=-(d\ln T/dr)^{-1}$, respectively. At the $q=2$ RS, we have set $R_0/L_{Ti}=R_0/L_{Te}=6.0$, $R_0/L_{ni}=R_0/L_{ne}=1.9$. The resolution is chosen as (400×100×32) grid points in ($r, \theta, \xi$) coordinates in order to ensure that the resolution on the poloidal section satisfies $r\Delta\theta \approx \Delta r \approx \rho_i$, where $\xi$ refers to toroidal direction of a tokamak. A periodic boundary condition is used in the toroidal direction in GTC, while linear boundary conditions are employed in the extrapolation and the finite difference region in the radial direction. One hundred ions and electrons cells are loaded in simulations in order to obtain sufficiently low particle noise. It is worth noting that simulating the multiscale interactions in this study requires a significant amount of computational resources, especially if more cells are included. While indeed increasing the number of cells can improve the accuracy of the simulations, such treatment is beyond the present computational capacity. On the other hand, extremely fewer cells would prevent us from accurately capturing the real physical processes, as the level of particle noise would be comparable to those of the perturbed quantities. Furthermore, since high-frequency modes with $\omega_H = k_{\parallel}k_{\perp}(\lambda_D/\rho_i)\omega_{pe}$ might be excited under DKE description, we have limit the time step to $\Delta t = 0.001R_0/C_s$, thus satisfying $\omega_H\Delta t < 1$ and numerical instabilities can be efficiently removed.

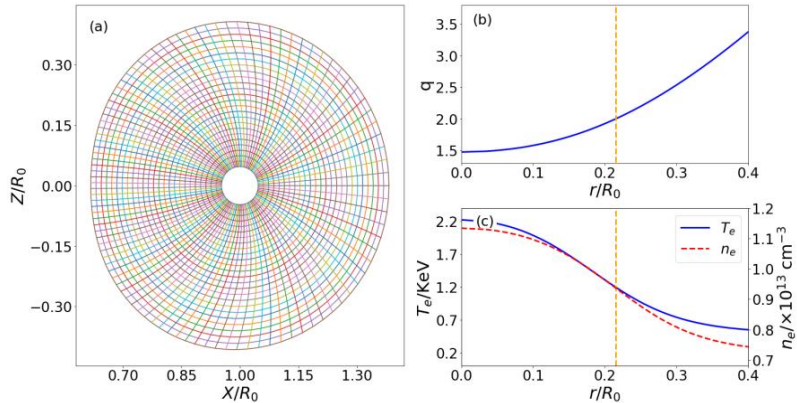

Fig. 1 (a) The grid mesh of the equilibrium, (b) the safety factor $q$ profile, (c) the temperature and density profile profiles utilized in the simulations. The vertical dashed line denotes the $q=2$ RS.

We added a magnetic island with $n/m = 1/2$, centered at $r = 0.5a$ corresponding to the location of $q = 2$ rational surface, whose width was adjusted by varying the coefficient manually. The magnetic island was implemented by a perturbed magnetic vector potential $\delta A_{I\parallel} = -\delta A_{\parallel 0} \times R_0 B_0 \cos(2\theta - \xi)$, where $R_0$ and $a$ are the major and minor radii of the tokamak, respectively. The helicity of the magnetic island was $m = 2$, $n = 1$. Similar to the definition of the magnetic island width in Ref. [29], the half width of the island $w$ is given here in terms of the amplitude of the parallel component of the vector potential $\delta A_{\parallel 0}$, the safety factor $q$, the magnetic shear $s=(r/q)(dq/dr)$, the magnetic field strength $B$, and the major radius $R_0$.

$$w^2 = \frac{2q\delta A_{\parallel 0} R_0}{Bs}$$

Based on the analysis presented in Ref. [40], we have carefully selected a specific set of toroidal mode numbers for our investigation, namely $n = 1, 9, 10$, and 11. It is worth noting that these distinct mode numbers have distinct physical interpretations: the $n = 1$ mode is associated with magnetic islands, while the $n = 9, 10$, and 11 modes are indicative of high-$n$ ITG instability modes.

*3.2. Effect of island width on turbulence and flows*
Figure 2 gives the time evolution of $\delta\phi^{10,20}$ (with the notation $\phi^{n,m}$ denoting the potential of a specified $m$ and $n$) under different MI width. The selection of such mode numbers is based on the fact that linear analysis has demonstrated that the $m/n=20/10$ ITG mode is most unstable in the present study [27]. Setting the electron temperature gradient as the same as the ion temperature gradient may excite both TEM and ETG modes. However, it is certain that revealing the electron modes requires a higher resolution in mode numbers thus only the ITG modes are retained in the work. The evolution of various modes calculated by GTC under similar parameters can be found in Figure 3 of Ref. [42]. A time window of 35 $R_0/C_s$ -50 $R_0/C_s$ in Fig. 2 was chosen for averaging. The horizontal lines do not represent the time period used for averaging in any arbitrary time window, but are used to show the values of different time windows during the saturation phase. The electrostatic potential of the ($m = 2, n = 1$) mode was diagnosed at the rational surface position with $q=2$, located at $r=0.5a$. It can be seen that in the absence of MI, the value of $\delta\phi^{10,20}$ is much lower than with a magnetic island. Moreover, the mean amplitude of $\delta\phi^{10,20}$ becomes larger when increasing the island width, as indicated by the horizontal lines. It is noticed that the turbulence driven zonal flows are the key saturation mechanism in our case thus it is strongly suggested that the coupling between the MI and turbulent flows is increased in the presence of a static magnetic island.

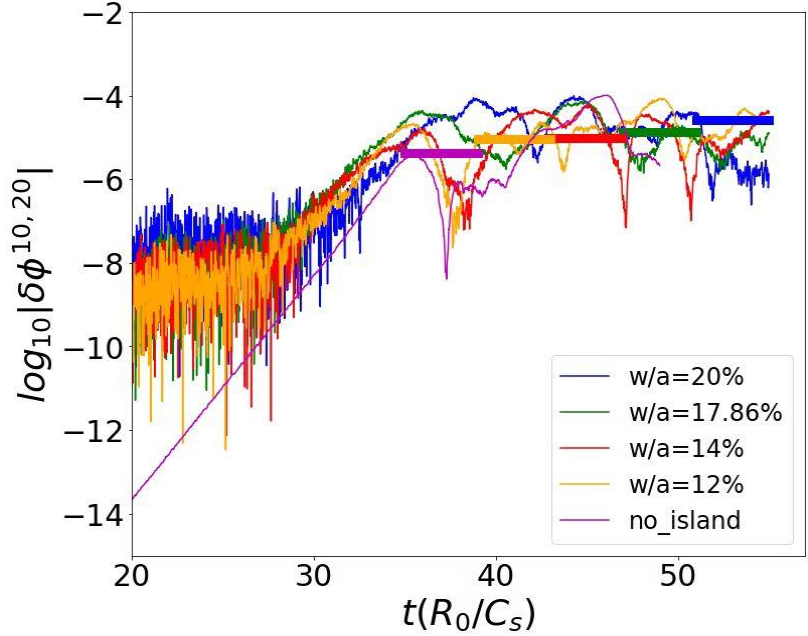

Figure 2 Time evolution of the amplitude of $\delta\phi^{10,20}$ under different MI width. A time window of 35 $R_0/C_s$ -50 $R_0/C_s$ was chosen for averaging.

Figure 3 shows the two-dimensional (2D) mode structure of magnetic island in terms of the perturbed electrostatic potential $\delta\phi^{1,2}$. Fig. 3(d) is that in the absence of MI while Fig. 3(a)-(c) are the corresponding structures with island width of $w/a$=14% $w/a$=17.86%, and $w/a$=20%, respectively. It is noted that $\delta\phi^{10,20}$ refers to high-$n$ modes that donotes the perturbations generated by ITG instability, and therefore is an ITG mode. Meanwhile, $\delta\phi^{1,2}$ corresponds to the modes associated with magnetic islands, which are used to characterize the electrostatic perturbations consistent with the helical features of the magnetic islands as mentioned in section 3.1. It can be seen that in the nonlinear stage, since the poloidal and toroidal mode structure tend to follow with the topology of the magnetic island, the structure of $\delta\phi^{1,2}$ is localized around the MIs, forming a large-scale vortex structure that elongated in the radial direction. Meanwhile, for the case without magnetic islands, $\delta\phi^{1,2}$ only appears as discrete small-scale structures in the radial and poloidal direction, as shown by Fig. 3(d). The formation of such kind of large-scale vortex structure (vortex flow) is generally considered to be similar to the generation mechanism of ZF, that is, when there are magnetic islands, the motions of ions and electrons along the magnetic lines cannot balance each other, and this asymmetry leads to the same excitation of large-scale structures which will be coupled with the magnetic island structures [43]. Obviously, from our results, the larger the magnetic island, the stronger the imbalance force is, and finally results in a more prevailing vortex flow.

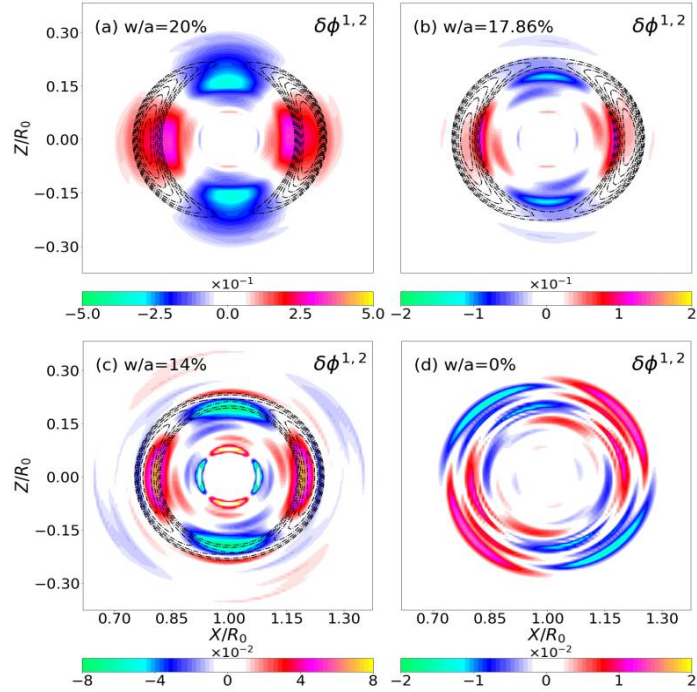

Fig. 3 Structure of the perturbed potential component $\delta\phi^{1,2}$ on the poloidal plane. (a) w/a=20% at t=47 $R_0/C_s$, (b) w/a=17.86% at t=49.6 $R_0/C_s$, (c) w/a=14% at 47.2 $R_0/C_s$ and (d) without magnetic island at 52 $R_0/C_s$. the black dashed lines represents the magnetic island structure. The time points are selected to ensure that the corresponding perturbation is at a maximum value.

Previous numerical simulations have found that the electrostatic vortex mode in the magnetic island will generate a strong $E \times B$ shear flow at the boundary of the magnetic island [30], thereby suppressing the drift wave instability and blocking the diffusion of external turbulence into the magnetic island. Here we define the helical coordinate and $E \times B$ flow as [40]:

$$\xi = m\theta - n\zeta, \quad (14)$$

$$v_\xi = -\frac{\nabla \delta\phi^{n>0} \times \boldsymbol{B}}{B_0^2} \cdot \boldsymbol{e}_\xi. \quad (15)$$

Here $\xi$ is the helical angle and $\delta\phi$ only contains the $n>0$ component of the potential fluctuation as the mode with $n=0$ is the conventional zonal flow which differs with the helical flow. The average spiral flow $\langle v_\xi \rangle_t$ was estimated during the simulation time interval $t = 35\,R_0/C_S$ to $t = 45\,R_0/C_S$. Figure 4 shows the typical profiles of the mean ion density, helical flow and 2D structure of the flow fluctuation in the presence of MIs, where Fig. 4(a)-(c) are the case with $w/a$=14% while (d)-(f) are that for $w/a$=8%. The profile flattening effect is clearly observed at $t = 39\,R_0/C_S$ which is enhanced by increasing the island width. We take into account the flattening effect within the magnetic island in all simulations. Only after the pressure is flattened within the island do we consider the nonlinear coupling process between the magnetic island and turbulence. Furthermore, the mean helical flow at the island boundary not only has a large amplitude, but also has a strong shear as the MI width becomes larger, which can be concluded by the comparison between Fig. 4(b) and 4(e). The poloidal structure of helical flow is mainly distributed along the boundary of the magnetic island. This shear flow is mainly generated by the (2, 1) vortex mode near O-point, while at the reconnection site, the high-$n$ structure of ITG turbulence and the (2, 1) structure corresponding to the magnetic islands exist simultaneously, as shown by Fig. 4(c). At the boundary of magnetic islands, the helical flow exhibits maximum shear or slope, with the most pronounced helical flow occurring at the reconnection point or magnetic island boundary, regardless of island size. In contrast, the helical flow at the O-point of the magnetic island is found to be the weakest. It can be inferred from Fig. 4(c) or 4(f) that the maximum gradient of the potential fluctuation, i.e., the radial electric fields will reach maximum just outside the island region, leading the strongest drive force for the zonal flows in terms of Reynolds stresses [44] as shown in Fig. 4(b) or 4(e). However, it should be noted that the maximum amplitudes of ZFs locate outside the island due to the suppression of turbulence by MIs around them. Furthermore, it is also found that the time required to fully flatten the density profile inside the magnetic islands is computationally long for small island width and amplitude. Small magnetic islands are also characterized by helical flow and shear near the separatrix, but the shear is relatively small which is suggested to be due to the fact that the amplitude of the vortex modes under the small MIs are relatively low. In our simulations, the low-$n$ components in the electrostatic potential are generally attributed to the vertex flow induced by magnetic islands, while the high-$n$ modes and the helical flow are considered to be caused by both MI and turbulence.

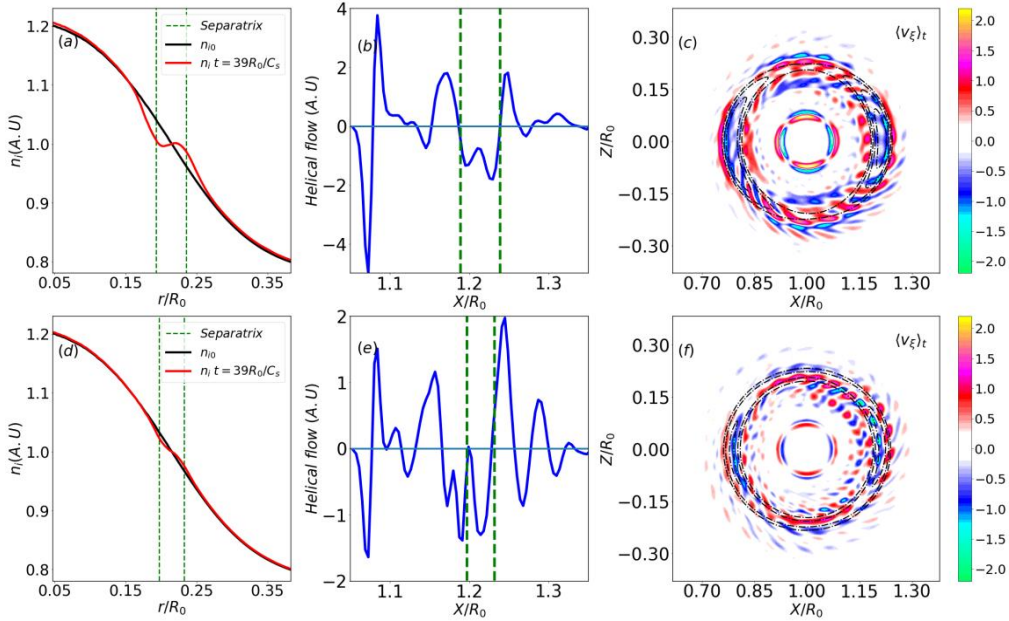

Fig. 4 Radial profiles of ion density and mean helical flow as well as the 2D structure of the flow perturbation at $t=$ . (a)-(c) are the case with $w/a$=14% while (d-f) are that for $w/a$=8%. The magnetic island structure is illustrated with the black dashed lines. The blue dotted line, the back line, and the red line denotes the separatrix of magnetic island, the initial ion density profile as well as the fattened ion density profile at t=39$R_0/C_s$.

The effect of island width on the amplitudes of zonal component is shown in Fig. 5(a). The diagnostics position is located at the $r/a$=0.5, and a time window of 35 $R_0/C_s$ -50 $R_0/C_s$ was chosen for averaging (except for the case of $w/a$=20% due to its relatively short time to reach the quasi-steady state). It is found that the zonal field, denoted by the root-mean-square (RMS) of the $n$=0 component, is increased with the MI width under large magnetic islands. The finding is different with previous studies which suggest that the tearing modes would suppress both turbulence and zonal flows [16]. We claim that the differences might strongly depend on the simulation models (fluid or gyrokinetic), methods (local or global) and equilibrium physical parameters.

This is easily to be understood as the linear and nonlinear properties of ITG turbulence strongly depends on the models and physical parameters, for instance, the ITGs are known to be subjected to the finite-β stabilization whereas we have assumed pure electrostatic ITG turbulence in this work. The geometrical and electromagnetic effects will be important aspects to be resolved in the forthcoming works. The effect of island width on zonal flows is depicted in Fig. 5(b), where it is shown that the scaling of ZF amplitude with MI width generally exhibits a nonmonotonic relationship, i.e., the ZF is suppressed by MI for the case of moderate widths as demonstrated by lots of simulation works, however, it is enhanced by the large magnetic island, as shown by Fig. 5(b). The case of $w/a$=8% is similar to that of $w/a$=12% as both of the the amplitudes of the MIs are relatively small, hence their effect on the turbulence and helical flows can not be clearly distinguished. The results can also be inferred from Fig. 5(b). The values and errorbars are extracted from the time ranges indicated by the lines by mean values and standard deviations in Fig. 5(a). The nonmonotonic relations among turbulence, GAM and island width were also partially demonstrated in the experiments on the HL-2A tokamak [25].

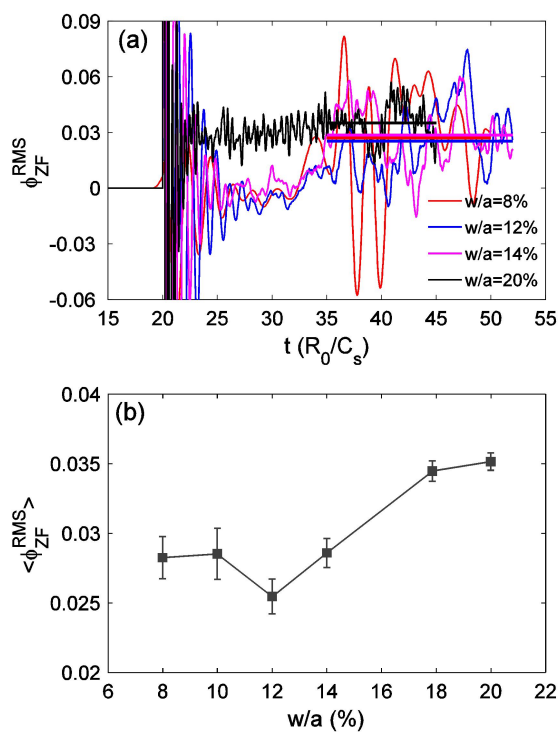

Fig. 5 (a) time evolution of zonal field for three MI width of $w/a$=8%, $w/a$=12%, $w/a$=14% and The structure of the ITG turbulence under different magnetic island width is illustrated in Fig. 6, which gives the full electrostatic potential structures of high-$n$ modes(with $n > 1$), were obtained at a specific time point, namely, $t = 46\ R_0/C_s$. It can be seen that the magnetic island has a clear impact on the structure in the nonlinear stage of ITG turbulence. The turbulence tends to reorganize itself around the island, resulting in a significant reduction in its amplitude when the island size is large. It can be seen that the ITG mode deviates from the original ballooning mode structure, and gradually converges around the X-point on both sides. In the fully developed nonlinear stage, the ITG mode within the magnetic island has almost completely disappeared, while more pronounced ITG turbulence emerges near the X-point on both sides of the separatrix. For instance, in Fig. 6(a), when $w/a$=20%, a clear ITG turbulence structure is present. As the magnetic island width decreases, the enhancement effect at the X-point weakens, and the mode structure around the O-point becomes more prominent, as seen in Fig. 6(d). When the magnetic island is small enough, an obvious mode ITG turbulence structure will appear at or near the O-point, however, the enhancement of the mode structure at the reconnection site can hardly be seen. Such results have suggested that the inhibitory effect of the MIs on ITG turbulence is very weak under the cases of small magnetic islands. The suppression of the ballooning structure by the magnetic island can be regarded as the flattening of the internal pressure of the magnetic island hence the driving force of the ITG turbulence becomes smaller.

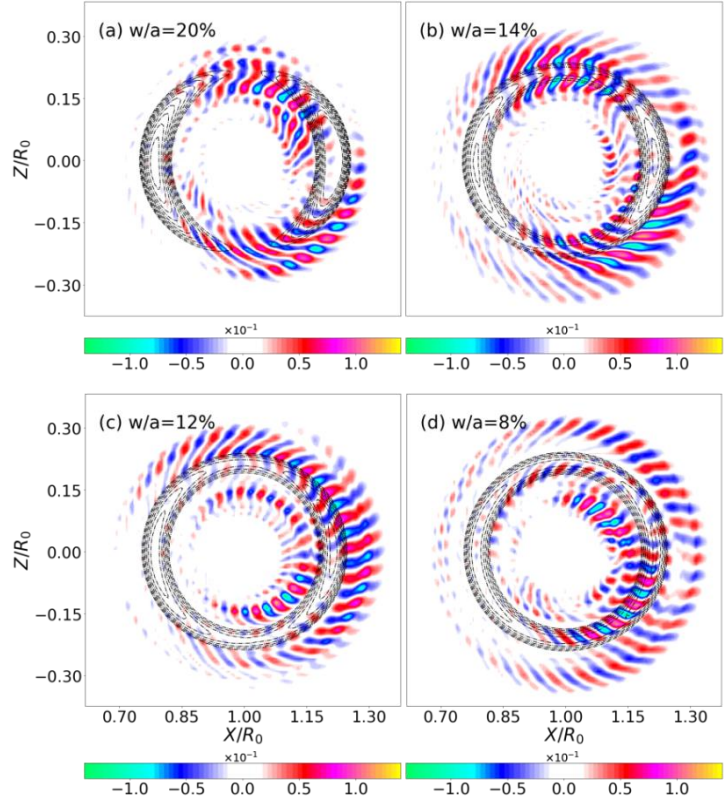

Fig. 6 The ITG turbulence structure under different island width: (a) *w/a*=20%, (b) *w/a*=14%, (c) *w/a*=10% and (d) *w/a*=8%. The black dashed lines represents the magnetic island structure.

In addition to the effects of magnetic island on the structure of both flows and turbulence, the influence of MIs on transport is also an important topic which will give a more clear understanding of the cross-scale interactions between magnetic islands and turbulence. The transport in the system of ITG turbulence in the presence of magnetic can be defined as follows:

$$\begin{cases} \tilde{n}_i(\boldsymbol{r},t) = n_i(\boldsymbol{r},t) - \langle n_i(\boldsymbol{r},t) \rangle_t, \\ \tilde{n}_i^2(\boldsymbol{r}) = \langle \tilde{n}_i^2(\boldsymbol{r},t) \rangle_t. \end{cases} \quad (16)$$

where the intensity of turbulence is treated as the root-mean-square values of the ion density fluctuation, which is denoted by the difference between the total density $n_i(\boldsymbol{r},t)$ and the average ion density $\langle n_i(\boldsymbol{r},t) \rangle_t$ within a given time period. We denote $\tilde{n}_i(\boldsymbol{r})$ as the turbulence intensity in the absence of magnetic islands, and $\tilde{n}_i^{'2}(\boldsymbol{r})$ as the turbulence intensity in the presence of magnetic islands. The role of magnetic islands can be written as their difference in turbulence intensity:

$$\Delta \tilde{n}_i^2(\boldsymbol{r}) = \tilde{n}_i^{'2}(\boldsymbol{r}) - \tilde{n}_i^2(\boldsymbol{r}). \quad (17)$$

Besides, we define the particle diffusion coefficient $D$, ion and electron thermal conductivity $\chi_i$ and $\chi_e$ as:

$$\begin{cases} D = \dfrac{1}{n_{0i}\nabla n_{0i}} \int d\boldsymbol{v} v_r \delta f_i, \\ \chi_x = \dfrac{1}{n_{0x}\nabla T_{0x}} \int d\boldsymbol{v} \left(\dfrac{1}{2}m_x v^2 - \dfrac{3}{2}T_x\right) v_r \delta f_x. \end{cases} \quad (18)$$

Here $v_r = v_{E\times B} + v_{\delta B}$ consists of electric drift velocity and radial flow induced by the magnetic island in the radial direction. Figure 7 shows the results of the influence of magnetic islands on turbulent transport under different magnetic island sizes. The transport coefficients are normalized by the gyro-Bohm unit which is defined as $D^{GB} = \chi_i^{GB} = \rho_i^2 v_i/a$. The normalization of transport coefficients ($D^{GB}$) is fixed and not depend on a given reference flux surface. It also should be emphasized that $\Delta \tilde{n}_i^2(\mathbf{r})$ corresponds to time-propagated values averaged over the time range of t=45-55$R_0/C_s$ in Fig.7. The turbulent intensity is continuously reduced inside the island while stronger fluctuations appear outside the island and at X-point when the MI width becomes larger, which can be seen from Fig. 7(a1)-(a3). When the magnetic island is rather small, the turbulence enhancement at X-point becomes less obvious; nonetheless, the phenomenon of turbulence suppression inside the island still remains. In addition, Fig. 7(c-d) suggests that the transport is almost fully suppressed inside the island region together with enhanced turbulence transport near the X-point, which is consistent with the ITG turbulence structure shown previously. As the magnetic island is as small as 8%*a*, both the transport enhancement by X-point and the reduction by O-point disappear, as can be found in Fig. 7(d1)-(d3). Nevertheless, there is still an enhancement of turbulent transport near the separatrix of magnetic island. These results might be closely related to the enhanced transport during resonant magnetic perturbations which induces island chains in the edge region, thus providing a transport channel for the particle transport [45]. Through our calculations, we have observed that even when *w*=9%*a*, the transport enhancement at the X-point remains present. Hence, we have determined the minimum magnetic island size, i.e., *w*=8%*a*=12$\rho_i$, that leads to the observed transport enhancement at the X-point. This implies that when the MI size exceeds this threshold, the transport enhancement by the X-point effect can be observed, while it disappears when the MI size is below this threshold. Regarding the quantitative relationship between the reduction of magnetic island turbulence and the width *w* of the island, it remains challenging to establish a definitive dependence between them at present.

In general, it is noted that the presence of the magnetic island tends to stabilize the turbulence around the O-point, which is consistent with previous gyrokinetic simulations [30, 46, 47]. When the width of the magnetic island is large (w > 8%a ≈ $12\rho_i$), the suppression of turbulence inside the island by the external turbulence is more significant. This is mainly due to the flattening of the pressure gradient inside the island, which weakens the linearly driven instability. Very small islands (w ≈ $12\rho_i$) seem to have little effect on turbulence, while the effect of islands with w ≥ $22\rho_i$ is more significant. This result is qualitatively similar to the result reported in Ref. [30], and an obvious enhancement of turbulence can be observed around the X-point. This is mainly because, in the simulation, the energy flux is the same on both sides of the simulation box. When the island reduces the horizontal turbulence around the separating point, energy will flow through a path that does not pass near the O-point, leading to an increase in fluctuation around the X-point. When the magnetic island is of intermediate to small scales (w < $12\rho_i$), the impact of the island on turbulence is relatively small, and a significant transport enhancement begins to appear at the O-point due to the weakening of turbulence drive. Similarly, it is the increase in transport enhancement around the O-point caused by the weakening of turbulence that leads to the screening of transport enhancement around the X-point.

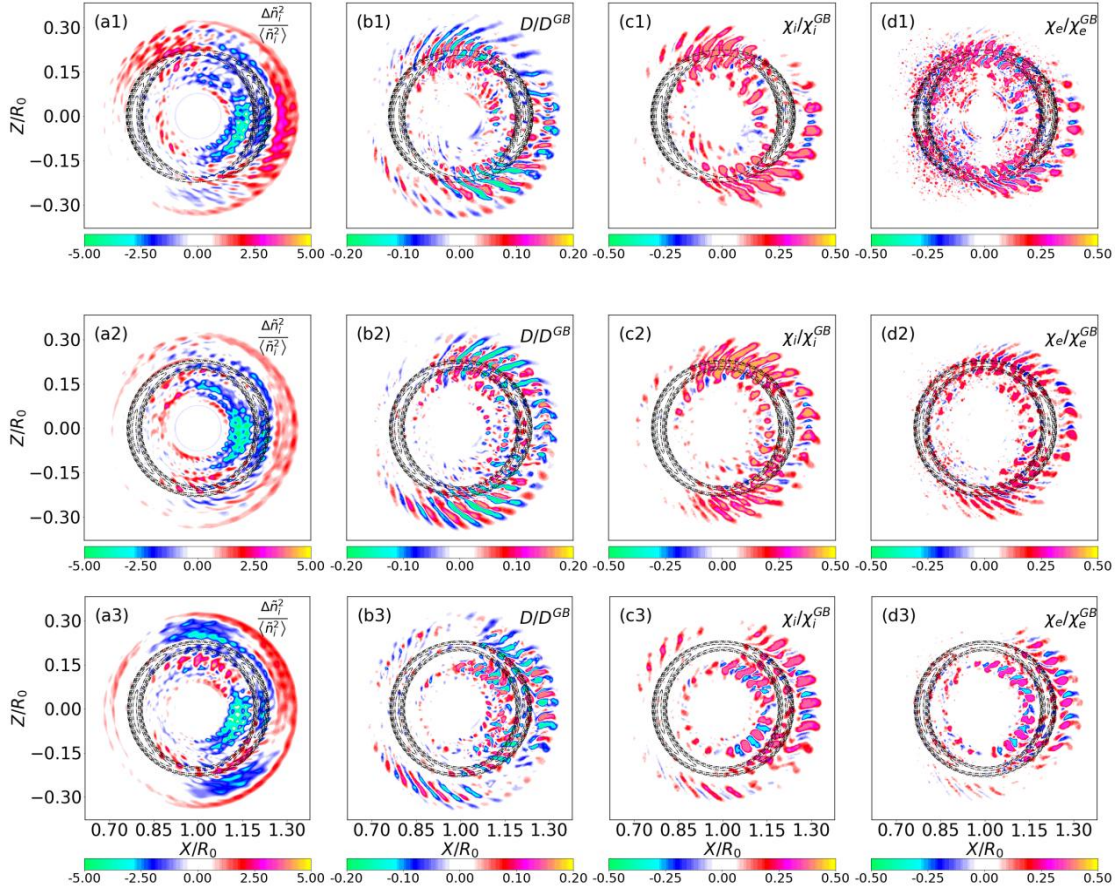

Fig. 7 The impact of MI width on turbulent intensity and transport in terms of relative amplitude, particle diffusion coefficient, ion and electron thermal conductive coefficients in gyro-Bohm units. (a1)-(d1): $w/a$=14%, (a2)-(d2): $w/a$=10%, (a3)-(d3): $w/a$=8%. The magnetic island structure is illustrated with the black dashed lines.

## 4. Conclusion and discussion

Global nonlinear simulations of the influence of magnetic island width on ion temperature gradient driven turbulence are carried out using the electrostatic gyrokinetic equation incorporated with static $m/n$=2/1 magnetic perturbation. The self-consistent evolutions of both the (2, 1) mode and ballooning parity microturbulence represented by ITG are achieved by introducing the flattening effect of the magnetic island on profiles. Results have shown that the amplitude of the high-$n$ mode increases with magnetic island width, indicating that the coupling between the magnetic island and the turbulence is stronger with a larger island size. The large-scale electrostatic structure, i.e., the so-called vortex flow exists around the island which drives a strong $E \times B$ shear flow at the boundary of the magnetic island whose amplitude and shear is increased with island width. Besides, the turbulence is enhanced at X-point whereas reduced at the O-point side once the island is large enough. Such effect is weakened as the island becomes smaller and almost disappears when $w/a$=8%, however, the turbulence is still influenced at the island separatrix. The turbulence transport is also largely affected by the magnetic island, showing that it decreases inside the island whereas increases outside it close to the X-point, which is in qualitatively consistent with recent experimental findings [48-50]. The presence of macroscopic magnetic islands usually has deleterious effects on the confinement and performance of tokamak plasmas [51, 52]. However, experimental observations have shown that depending on the strength of perturbations and plasma responses, magnetic islands can also enhance thermal confinement and reduce turbulent transport levels. In particular, their beneficial role in forming internal transport barriers (ITBs) has been demonstrated in experiments [18, 48]. In the HL-2A tokamak, it has been observed that the perpendicular flow is very small at the center of the island near the O-point and strongly enhanced around the boundary of the island and the density fluctuations inside the island are generally weakened. This suggests that magnetic islands can suppress local turbulence [49]. In KSTAR, two-dimensional profiles of electron temperature and poloidal flow measurements have shown that magnetic islands can act as either barriers or fast channels of electron heat transport. When the poloidal flow is perturbed to have a vortex structure, the magnetic island acts more like a barrier with respect to electron heat transport. The positive flow shear in the inner region suppresses electron temperature turbulence near the O-point, and only a narrow region close to the X-point shows significant electron temperature turbulence levels [50]. Therefore, our simulation results are in well agreement with recent experimental results.

The suppression effect by magnetic island on the global transport properties generally vanishes as long as the island is sufficiently small, however, it still exists near the separatrix of the island. Moreover, as mentioned earlier, the influence of polarization current on the interplay between MIs and turbulence cannot be disregarded. Nonetheless, it is challenging to distinguish the specific contribution of the polarization current from other nonlinear interactions in the framework of gyrokinetic study. Therefore, studying the polarization current effects during the interaction between turbulence and magnetic islands is currently not feasible and will be carried out in the future.

**Acknowledgments**

This work was partly supported by National Key R&D Program of China (Grant Nos. 2018YFE0303102, 2019YFE03020004 and 2022YFE03060000), National Natural Science Foundation of China (Grant Nos. 11905109, 11947238, 12075079 and 12125502), the U.S. Department of Energy, Office of Science, Office of Advanced Scientific Computing Research and Office of Fusion Energy Sciences, and the Scientific Discovery through Advanced Computing (SciDAC) program under Award No. DE-SC0018270 (SciDAC ISEP Center), and Sichuan Science and Technology Program (Grant No. 2021JDJQ0029). We also acknowledge the Center for Computational Science and Engineering of Southern University of Science and Technology for providing computational resources.